\documentclass[12pt,aps,onecolumn,nobibnotes]{revtex4}

\usepackage{graphicx}%
\usepackage{dcolumn}
\usepackage{amsmath}
\usepackage{amssymb}

\begin{document}

\title[Regularization by discrete symmetries]{%
REGULARIZATION OF NON-LINEAR SPINOR FIELD MODELS 
BY DISCRETE SYMMETRIES}
\author{\underline{Bertfried Fauser}}
\email{Bertfried.Fauser@uni-konstanz.de}
\author{Heinz Dehnen} 
\email{Heinz.Dehnen@uni-konstanz.de}
\affiliation{%
Fachbereich Physik\\
Universit\"at Konstanz, Fach M 678\\ 
78457 Konstanz, Germany\\ 
}
\begin{abstract} 
We generalize a regularization method of Stumpf \cite{Stumpf84} in
the case of non-linear spinor field models to fourth order theories 
and to non-scalar interactions. The involved discrete symmetries can 
be connected with ${\sf C}$, ${\sf P}$, ${\sf T}$ transformations.
\end{abstract} 
\maketitle 

\vspace{-1.0truecm}
\section{Introduction}
Beside being obviously necessary for obtaining results,
regularization has to be included into the definition
of a theory. Different methods of regularization may lead to
different outcomes. Renormalization is a special kind
of regularization for singularities which occur only perturbatively. 
Since non-linear spinor field models are non-renormalizable, 
we are exclusively interested in a regularization process. 
 
Using {\it higher order derivatives\/}, one can make a theory
more regular or even fully regular. As can be seen e.g. by power
counting arguments. However, the interpretation of the involved
fields becomes a difficult task, as also canonical quantization
is no longer obvious (spin-statistic theorem). We will stay with 
quantization of first order fields and calculate the commutators 
of the higher order fields afterwards.

We extend the canonically quantized two-field theory with
scalar interaction of Stumpf \cite{Stumpf84} to a fully regular
four-field theory with non-trivial $V$-$A$ interaction. 
\section{Stumpf 2-Field Theory}
We define a non-linear Heisenberg \cite{Hei} or Nambu--Jona-Lasinio 
\cite{NamJon} like first order spinor field model by
\[
(i\gamma^\mu\partial_\mu - m)_{IJ} \psi_J =
g\,V_{IJKL} \,:\, \psi_J \psi_K \psi_L \,:\,
\]
using a compactified index notation
\begin{eqnarray}
\psi_I \,=\, \psi_{i\lambda}:= \left\{
\begin{array}{cl}
\psi_i \,=\,
\psi({\bf x},t)_{\alpha,\ldots,\beta} & \mbox{if~~} \lambda=0 \\
\psi_i^\dagger \,=\,
\psi^\dagger({\bf x},t)_{\alpha,\ldots,\beta} & \mbox{if~~} \lambda=1
\end{array}\right.
\end{eqnarray}
where $\alpha,\ldots,\beta$ are some algebraic degrees of freedom
and $V_{IJKL}$ is a constant vertex function antisymmetric in the
last three indices $V_{IJKL} = V_{I[JKL]_{as}}$ and $g$ is the
coupling constant. 

The Stumpf model of regularization is restricted to the massless 
theory using scalar interaction. Parity ${\sf P}$ is defined as
$({\vec {\bf r}},t) \mapsto (-\vec{ {\bf r}},t)$
and $\psi \mapsto i\gamma_0 \psi$ yielding 
\[
\psi^\prime_i({\vec {\bf r}^\prime},t) :=
{\sf P} \psi_i({\vec {\bf r}^\prime},t) \,=\,
i\gamma_0 \psi_i({\vec {\bf r}},t),\quad\quad
{\vec r^\prime} \,=\, - {\vec r}.
\]
Lagrangian and Hamiltonian are given as usual. One has the common 
invariance $L[\psi^\prime] = L[\psi]$ and $H[\psi^\prime] = H[\psi]$.
 
Let us introduce an auxiliary field which is only the
spin-parity transform, but {\it not\/} the space-parity transform as 
\begin{eqnarray}
\xi_i \,:=\, \psi^\prime_i({\vec {\bf r}},t)
\,=\, \psi^\prime_i({-\vec {\bf r}^\prime},t)
\,=\, i\gamma_0 \psi_i({\vec {\bf r}},t).
\end{eqnarray}
This is sufficient to prove an equivalence theorem 
$
H[\Psi] \,\equiv\, H[\Psi,\xi]\vert_{\xi=i\gamma_0\Psi},
$
where the two-field Hamiltonian reads
\[
H[\Psi,\xi] \!\!:=\!\! \frac{i}{4}\int[
\bar{\Psi}(t,\vec{\bf r})\vec{\gamma}\cdot\vec{\nabla}\Psi(t,\vec{\bf r})
-\vec{\nabla}\bar{\Psi}(t,\vec{\bf r})\cdot\vec{\gamma}\Psi(t,\vec{\bf r})
]dr^3
\]
\[
-\frac{i}{4}\int[\bar{\xi}(t,\vec{\bf r})\vec{\gamma}\cdot\vec{\nabla}
\xi(t,\vec{\bf r})
-\vec{\nabla}\bar{\xi}(t,\vec{\bf r})\cdot\vec{\gamma}\xi(t,\vec{\bf r})
]dr^3
\]
\[
+\frac{1}{4}g\int\{[\bar{\Psi}(t,\vec{\bf r})+\bar{\xi}(t,\vec{\bf r})]
[\Psi(t,\vec{\bf r})+\xi(t,\vec{\bf r})]\}^2dr^3
\]     
The proof needs scalar interaction \cite{Stumpf84}. Transforming back 
to the Lagrangian picture results in two inequivalent Lagrangians
\[
\begin{array}{ccc}
 H[\Psi] & = & H[\Psi,\xi]\vert_{\xi=i\gamma_0\Psi}\\[2ex]
 \Updownarrow& &\Updownarrow\\[2ex]
 L[\Psi] & \not= & L[\Psi,\xi]\vert_{\xi=i\gamma_0\Psi}
 \end{array}
\]
This should be not a miracle, since we let open the constraint
that the second field $\xi$ is the parity transformed field $\psi$.
Moreover, quantization breaks the equivalence. 
Introduce $\varphi_1 \equiv \psi$, $\varphi_2 \equiv \xi$ for convenience
and the corresponding canonical momenta $\Pi_i$. We contrast the
one-field quantization with commutation relations
given as 
\[
\Pi(t,\vec{\bf r}):=i\Psi^\dagger(t,\vec{\bf r})=
\frac{\delta {\cal L}[\Psi]}{\delta \partial_t\Psi},
\]
\[
\{\Pi(t,\vec{\bf r}^\prime),\Psi(t,\vec{\bf r})\}_+=\delta
(\vec{\bf r}^\prime-\vec{\bf r})
\]
with the {\it two-field case\/} where we obtain from the two
field  Lagrangian and canonical quantization
\[
\Xi_i(t,\vec{\bf r}):=\frac{\delta {\cal
L}[\varphi_1,\varphi_2]}{\delta \partial_t\varphi_i}=
\frac{i}{\lambda_i}\varphi_i^\dagger(t,\vec{\bf r}),
\]  \[
\{\varphi_j^\dagger(t,\vec{\bf r}^\prime),\varphi_i(t,\vec{\bf r})\}_+
=
\lambda_j\delta_{ji}\delta(\vec{\bf r}^\prime-\vec{\bf r}) \,.
\]  
These two different quantization schemes break explicitely the equivalence
of the theories. One observes that 
$\varphi_1(t,\vec{\bf r}) \not= i\gamma^0\varphi_2(t,\vec{\bf r})$ 
and the auxiliary field particle number operators $N[\varphi_i]$ do not 
commute with the two-field Hamiltonian $H[\varphi_1,\varphi_2]$. 
However, their sum does commute and one introduces the physical 
field $\Psi = \sum \varphi_i$ while extending the theory as a great 
canonical ensemble
\[
K[\varphi_1,\varphi_2] := H[\varphi_1,\varphi_2]+\mu_1 N[\varphi_1]
 +\mu_2 N[\varphi_2] \, .  
\]
with $\mu_i$ as chemical potentials. Finally one obtains the 
{\it second order\/} field equations
\[
\Pi_{i=1}^2(i\gamma^\mu\partial_\mu-\mu_i)\Psi(t,\vec{\bf r})
\,=\,
(i\gamma^\mu\partial_\mu-\mu_1)(i\gamma^\nu\partial_\nu-\mu_2)
\Psi(t,\vec{\bf r}) = \frac{1}{4}g\left\{
\bar{\Psi}(t,\vec{\bf r})\Psi(t,\vec{\bf r})\right\}\Psi(t,\vec{\bf r}) 
\]
with the commutation relations and masses (chemical potentials)
\[
\{\bar{\Psi}(t,\vec{\bf r}^\prime),\Psi(t,\vec{\bf r})\}_+ = 0
\quad\quad\quad
\lambda_i = \frac{1}{\mu_i-\mu_j},\quad i\not= j,\quad \mu_i\not=\mu_j.
\] 
The expression for the $\lambda_i$ as function of the $\mu_i$
follows also in the reversed argumentation by fraction into parts
the Fourier transformed equation \cite{StuBor}.
\section{Extension to 4 Auxiliary Fields}
The results presented in this section have been obtained using CLIFFORD
a Maple V package for Clifford algebras developed by R. Ab{\l}amowicz
\cite{CLIFFORD}. Even using computer algebra, we have not been able to 
exhaust all possible interactions, but used a reduced picture, where we 
allowed scalar $S$, vectorial $V$, axial-vectorial $A$, and pseudo
scalar $P$, interactions only.

To extend the theory \cite{Fau-thesis}, we need to examine the mechanism
behind the proof that $H[\Psi] \equiv H[\varphi_1,\ldots,\varphi_n]$ with 
$n$-auxiliary fields. It can be shown that 
\[
\sum_{i,j \atop i\not= j} \varphi_i M_{ij} \varphi_j = 0
\]
is sufficient, where $\varphi_i := {\sf T}_i \psi$ and the ${\sf T}_i$ 
form a discrete group with ${\sf T}_i^2 = \pm 1$.

To obtain 4 auxiliary fields we need the (pseudo) quaternion group $G$ of two 
generators ${\sf T}_1$, ${\sf T}_2$. $G$ is spanned by $G\, =\, <{\sf T}_0 = Id,
{\sf T}_1,{\sf T}_2, {\sf T}_3={\sf T}_1{\sf T}_2>$. 
Defining $\varphi_i = {\sf T}_i \psi$  we have to investigate 5 different 
signatures which remain after utilization of permutation
symmetry on the labels:
\[
\{ {\sf T}_0^2,{\sf T}_1^2,{\sf T}_2^2,{\sf T}_3^2 \}
\!\in\! (\pm 1,\pm 1, \pm 1, \pm1); 
\quad\quad
sig \in (+4,+2,0,-2,-4).
\]

Examining the different signatures to obtain 4 eigenvectors,
suitable normalized and fulfilling the desired equation to 
perform the equivalence proof $H[\Psi] \equiv H[\varphi_1,
\ldots,\varphi_3]$ we find that only sig $=0$ yields a reasonable 
result. The corresponding Hamiltonian is calculated to be
 \[
H[\psi_i]=\frac{i}{2}\int\sum_{j=1}^4
\frac{1}{\lambda_j}\Big[\bar{\psi}_j\gamma^k\partial_k \psi_j
               -\partial_k\bar{\psi}_j\gamma^k\psi_j\Big]dr^3
\]
\[
+\frac{g}{16}\int\Big[
\sum_{j,k,l,m=1}^4 \bar{\psi}_j\gamma^\mu\psi_k
                   \bar{\psi}_l\gamma_\mu\psi_m
\]
\[
+\mbox{~~}
\sum_{j,k,l,m=1}^4 \bar{\psi}_j\gamma^\mu\gamma^5\psi_k
                   \bar{\psi}_l\gamma_\mu\gamma^5\psi_m\Big]dr^3
\]

This regular case has
$\lambda_1=\lambda_2=-\lambda_3=-\lambda_4=8$.
The canonical commutation relation yields rewritten in
$\Psi^\dagger,\Psi$ once more 
$\{\psi_i^\dagger,\psi_j\}_+=\lambda_i\delta_{ij}$ where two
$\lambda_i$'s are negative. We define the sum-field
$\Psi=\sum_{i=1}^4\psi_i$ with 
$\lambda_i=\prod_{j,i\not= j}1/(\mu_i-\mu_j)$.
Calculating the corresponding field equation yields
\[
\Big[\prod_{i=1}^4(i\gamma^\mu\partial_\mu-\mu_i)\Big]\Psi=
\frac{g}{16}\Big[
\bar{\Psi}\gamma^\mu\Psi\gamma_\mu\Psi+
\bar{\Psi}\gamma^\mu\gamma^5\Psi\gamma_\mu\gamma^5\Psi\Big]
\]
which is of {\it fourth order} and obeys the Heisenberg
non-canonical quantization rule for the sum-field. 
\[
\{\Psi,\Psi\}_+=0.
\]
Identifying the discrete symmetries with ${\sf C}$, ${\sf P}$, 
${\sf T}$, as already done in the two-field case with ${\sf P}$,
the 4 auxiliary subfields are given up to a phase by
$
\varphi_0 = {\sf P}\psi
$, 
$
\varphi_1 = {\sf CP}\psi
$,
$
\varphi_2 = {\sf CPT}\psi
$, 
$
\varphi_3 = {\sf TP}\psi
$.
The theory with 4 auxiliary fields leads to a 
canonically quantized sub-field theory which is fully regular
and of forth order in the physical sum-field $\Psi$.
\section{Discussion}
Our constructive model supports the approach given by Stumpf 
postulating higher order field equations with non-scalar vertex 
and factorizing them into canonically quantized non-particle 
auxiliary fields. The benefits of this method over the pragmatic
introduction of auxiliary fields in \cite{Stumpf84,StuBor} are:

We get a dynamical regular theory by constructive methods. Many 
calculations have shown that no divergent integrals occur \cite{StuBor}.

The `masses' have to be addressed as chemical potentials, even if they 
occur in Dirac spinor theory at the place where mass-terms are convenient.
This is important, sine at least one of these chemical potentials has to 
be negative if a regularization shall take place. This is a reasonable 
situation only if the $\mu_i$ are chemical potentials not masses.

The auxiliary or sub-fermion fields possess not a proper particle 
interpretation. This is not an artifact, but a main ingredience in 
the Heisenberg-Stumpf theory of bound states which constitute the
physical observable particles like quarks and leptons, but also bosons 
like gluons, the photon and electro weak W and Z bosons \cite{StuBor,Hei}.

Viewed in a more conventional setting \cite{NamJon}, the additional
parameters are exactly the counter-terms in a Pauli-Villars regularization
\cite{StuBor,Fau-thesis}. 

We derive from the canonical quantization of the sub-fields 
a non-canonical quantization of the physical sum-field which 
is of Heisenberg type. This usually {\it ad hoc\/} quantization rule 
is known to be of regular nature.

Most interesting is that the $\varphi_i$ fields are related by the 
discrete subgroup of the ${\bf pin(1,3)}$ group. The relation of this 
fact to the framework of Clifford analysis will be shown elsewhere.
\section{Acknowledgement}
Computations have been performed using CLIFFORD by R. Ab{\l}amowicz
\cite{CLIFFORD}. A travel grant for BF. from the DFG is greatfully 
acknowledged.

\end{document}